\documentclass[amsmath,superscriptaddress,amssymb]{revtex4}

\usepackage{epsfig}

\def\Journal#1#2#3#4{{#1} {\bf #2}, #3 (#4)}

\def\NPB{{\em Nucl. Phys.} B}
\def\PLB{{\em Phys. Lett.}  B}
\def\PRL{\em Phys. Rev. Lett.}
\def\PRD{{\em Phys. Rev.} D}
\def\PR{\em Phys. Rev.} 

\def\AP{\em Annalen der Physik}
\def\IJMPA{{\em Int. J. Mod. Phys.} A}
\def\IJMPD{{\em Int. J. Mod. Phys.} D}
\def\MPLA{{\em Mod. Phys. Lett.} A}
\def\CQG{\em Class. Quantum Grav.}
\def\AnJ{\em Astron. J.}
\def\AJ{\em Astrophys. J.}
\def\AJL{\em Astrophys. J. Lett.}
\def\N{\em Nature}
\def\AA{\em Astron. Astrophys.}
\def\GRG{\em Gen. Rel. Grav.}
\def\ARNPS{\em Ann. Rev. Nucl. Part. Sci.}
\def\ARAA{\em Annu. Rev. Astron. Astrophys.}
\def\AIHPA{{\em Ann. Inst. H. Poincar\'e} A}
\def\ASR{\em Adv. Space Res.}
\def\ESP{\em ESA Spec. Publ.}
\def\SPAWB{\em Sitz. Preuss. Akad. Wiss. Berlin}
\def\NGWG{\em Nach. Ges. Wiss. G\"ottingen}

\def\url#1{{\rm #1}}

\def\be{\begin{equation}}
\def\ee{\end{equation}}
\def\bea{\begin{eqnarray}}
\def\eea{\end{eqnarray}}

\def\g{g}

\def\munu{{\mu\nu}}
\def\h{h}                  
\def\R{R}                  
\def\E{E}                  

\def\PN{\Phi_N}
\def\PP{{\Phi_P}}
\def\Pst{\phi}

\def\x{x}                  
\def\t{t}                  
\def\s{s}                  
\def\xb{\mathbf{x}}       
\def\xbx{\xb_1}
\def\xbxm{{\xb_1^-}}
\def\xbxp{{\xb_1^+}}
\def\xby{{\xb_2}}
\def\r{r}                  
\def\rx{{\r_1}}
\def\ry{{\r_2}}
\def\rydd{{\ddot \r}_2}
\def\u{{u}}                
\def\ri{\rho}              
\def\th{\theta}            
\def\ang{\varphi}
\def\angxy{\phi}
\def\dang{{\dot \angxy}}
\def\k{k}                  
\def\ad{a}                 

\def\T{T}
\def\pro{\pi}              

\def\td{{\cal T}}

\def\annfactor{{\cal A}}
\def\sec{{\rm sec}}
\def\ann{{\rm ann}}
\def\zg{\zeta}
\def\prec{\varpi}
\def\ep{e}

\def\c{c}                  
\def\M{M}                  
\def\G{G}
\def\aP{a_{P}}
\def\H{\ell}
\def\aE{a_1}
\def\Ox{\Omega_1}
\def\tconj{t_\mathrm{conj}}

\def\stand#1{\left[#1\right]_\mathrm{GR}}
\def\unit#1{{\rm #1}}

\def\eprint#1{{\it Preprint} #1}
\def\etal{\textit{et al }}
\def\ibid{\textit{ibidem }}

\begin{document}

\title{Gravitation in the solar system and metric extensions of general relativity}

\author{Marc-Thierry Jaekel}
\address{Laboratoire de Physique Th\'eorique de l'Ecole Normale Sup\'{e}rieure, 
CNRS, UPMC, 24 rue Lhomond, F75231 Paris Cedex 05}

\author{Serge Reynaud}
\address{ Laboratoire Kastler Brossel, Universit\'{e} Pierre et
Marie Curie, case 74, CNRS, ENS, Campus Jussieu, F75252 Paris Cedex 05}

\begin{abstract}
Despite its remarkable agreement with gravity tests in the solar system, 
general relativity is challenged at larger scales. Large scale tests 
performed in the solar system by the Pioneer 10/11 probes have failed 
to confirm the expected gravity laws. We discuss here extensions of general 
relativity which preserve its geometrical nature while showing an ability 
to conciliate the Pioneer anomalies with other gravity tests.
Further testable phenomenological consequences are also briefly discussed.
\end{abstract}

\maketitle

\section{Introduction}

The regularly improving agreement of gravity tests performed in the solar system
with general relativity (GR) \cite{Will01,Fischbach98} has imposed the latter as 
the standard theory for gravitation. 
However, anomalies appear at very large scales, as in the rotation cuves of 
galaxies \cite{Aguirre,Riess} or in the relation betweeen redshifts and 
luminosities of some type II supernovae \cite{Perlmutter99apj,Perlmutter99prl}. 
They can be cured either by introducing "dark matter" and "dark energy" or 
by modifying the gravity law at large scales \cite{Sanders02,Lue04,Turner04}. 

Testing gravity laws at large scales is clearly of great interest in this context.
Such a test was performed in the extended mission of Pioneer 10/11 probes and
it failed to confirm the known laws of gravity \cite{Anderson98,Anderson02}.
The Doppler tracking data of the Pioneer probes exhibited deviations from the 
predictions of GR, the Doppler velocity residual varying linearly with time.
This "Pioneer anomaly" is thus naturally represented as an acceleration $\aP$ 
directed towards the Sun, with an approximately constant amplitude over a 
large range of heliocentric distances (AU $\equiv$ astronomical unit)
\bea
\label{Pioneer_acceleration}
\aP = (0.87 \pm 0.13) ~\unit{nm}~\unit{s}^{-2}\quad ,\quad 
20~\mathrm{AU}\le\r\le 70~\unit{AU}  &&
\eea
The anomaly adds to the puzzle of large scale gravity, as it may reveal an 
anomalous behaviour of gravity 
already at scales of the order of the size of the solar system.

Although a number of different mechanisms has been considered to this aim
\cite{Anderson02b,Anderson03,Nieto04,Turyshev04},
the anomaly has escaped up to now all attempts of explanation as a systematic
effect generated by the spacecraft itself or its environment \cite{Nieto05a,Nieto05}. 
This has motivated efforts devoted to recovering and reanalyzing Pioneer data 
over the whole duration of the Pioneer 10/11 missions \cite{Turyshev06}, 
as well as proposals for new space missions dedicated to the study of 
deep space gravity \cite{PAEM05}.

The present paper is devoted to the key issue of a confrontation of the 
Pioneer anomaly with theories of gravity.
We will show that a family of metric extensions of GR can be defined 
in a natural way and that it shows the capability of making the Pioneer anomaly 
compatible with other gravity tests. 
We will also briefly discuss further testable phenomenological consequences
of this framework.

\section{Gravity tests in the solar system}
\label{tests}

We first briefly review the agreement of GR with tests performed up to now
 in the solar system. 
Before discussing this question, one must first recall the two 
building blocks which found  GR as a gravitation theory. 
The first one, the equivalence principle, states the universality of free fall and gives GR its geometric nature.
Violations of the equivalence principle are usually looked for by following the difference $\eta$ 
in the free fall accelerations undergone by two test bodies having different
matter compositions. This parameter $\eta$ is constrained by modern experiments to remain extremally small, below the $10^{-12}$ level, so that the equivalence principle is one of the best tested properties of nature.
The validity of the equivalence principle has also been tested over a large range of distances, from the
millimeter \cite{Adelberger03}
to the sizes of Earth-Moon \cite{Williams96} or Sun-Mars orbit 
\cite{Hellings83,Anderson96}.
Obviously, the level of precision attained by tests precludes an explanation of the Pioneer anomalies as a violation of the equivalence principle, as  
those would occur at the  $10^{-3}$ level (at the largest explored distances).

The second basic property characterizes  GR as a field theory by prescribing a particular coupling  between
gravitation and its sources. This coupling is equivalent to the gravitation equations which allow one to determine the metric
tensor from the distribution of energy and momentum in space-time.
According to GR, the coupling between curvature
tensor and stress tensor takes a particularly simple form \cite{Einstein15,Hilbert,Einstein16}, as
one curvature tensor (Einstein curvature $\E_\munu$) is simply proportional to the stress tensor 
$\T_\munu$. This coupling then involves a single constant, the Newton gravitation 
constant $\G_N$ 
\bea
\label{GR_gravitation_law}
\E_\munu \equiv \R_\munu-{1\over2}\g_\munu \R={8\pi\G_N \over\c^4}\T_\munu
\eea
Such a relation may be seen as the simplest way of relating curvature and energy-momentum tensors. As a result of Bianchi identities,  $\E_\munu$ 
corresponds to the particular curvature tensor having a null covariant divergence, like the energy-momentum tensor. Whilst the first property follows from Riemannian geometry, the second one expresses the
conservation of energy and momentum and describes the geodetic motion undergone by freely falling  test bodies. One notes however that these properties do not by themselves imply Einstein-Hilbert choice.

In standard gravitation theory, the metric tensor in the solar system is deduced
by solving Einstein-Hilbert equation (\ref{GR_gravitation_law}) with the Sun as gravitation source. Rotation and non sphericity of the Sun may be discarded
for our present purpose, so that the gravity source 
is described as a point-like motionless body of mass $\M$ and
the metric may simply be written in terms of the Newton potential $\Pst$.
The solution is conveniently written in terms of spherical coordinates
($\c$ denotes light velocity, $\t$ and $\r$ the time and radius,
$\th$ and $\ang$ the colatitude and azimuth angles)
with the gauge convention of isotropic spatial coordinates 
\bea
\label{isotropic_metric}
d\s^2 &=& \g_{00} \c^2 d\t^2 + \g_{\r\r} \left( d\r^2  +
\r^2(d\th^2 + {\rm \sin}^2\th  d\ang^2) \right)\nonumber\\
\label{GR_solar_metric}
\g_{00} &=& 1+2\Pst+2\Pst^2+\ldots\quad, \quad
\g_{rr} = - 1+2\Pst+\ldots \nonumber\\
\Pst &\equiv& -{\G_N\M\over\c^2\r}, \qquad 
\left\vert\Pst\right\vert \ll 1
\eea
Then GR, or Einstein-Hilbert equations,
(\ref{GR_gravitation_law}) can be tested by comparing the gravitation effects entailed by solutions to this equation with those predicted by alternative theories. For this purpose, 
one embeds GR within a family of well defined and computable theories. This is usually done by confronting GR with the family of 
parametrized post-Newtonian (PPN) metric theories \cite{Eddington,Robertson,Schiff66,Nordtvedt68,WillNordtvedt72} 
\bea
\label{PPN_0}
\g_{00} = 1 + 2 \alpha \Pst + 2 \beta \Pst^2 + \ldots \quad,\quad 
\g_{\r\r} = -1 + 2 \gamma \Pst + \ldots &&
\eea
Among the parameters that can be considered, the constants $\alpha$, $\beta$ and $\gamma$ appear as the most relevent ones. The 
first one can be set to unity by redefining Newton constant $\G_N$,
GR corresponding to $\gamma=\beta=1$. The parameters  $\gamma$ or $\beta$ affect geodesic motions in a way which can be evaluated, leading to comparison with observations and providing  best fit values for  these parameters (\ref{PPN_0}). 

Experiments performed up to now have put
strict bounds on the potential deviations $\gamma-1$ and $\beta-1$ which have improved with time. From 
Doppler ranging on Viking probes in the vicinity of Mars
\cite{Hellings83} to deflection measurements using VLBI astrometry \cite{Shapiro04} or radar ranging on the Cassini probe \cite{Bertotti03},
possible values of $\gamma-1$ have decreased to less than a few $10^{-5}$. 
Precession of planet perihelions \cite{Talmadge88} and 
polarization by the Sun of the Moon orbit around the Earth 
\cite{LLR02} constrain linear superpositions 
of $\beta$ and $\gamma$ fixing  $\beta-1$ to be 
smaller than a few $10^{-4}$.

A complementary test of GR is to check the dependence of the Newton potential $\Pst$ 
(component $\g_{00}$ in (\ref{GR_solar_metric})) on the radial coordinate.
Modifications of Newton standard expression are usually parametrized 
in terms of an additional  Yukawa potential depending on two parameters, 
the range $\lambda$ 
and the amplitude $\alpha $ measured with respect to Newton potential $\Pst$
\bea
\label{Yukawa_perturbation}
\delta \Pst(\r) = \alpha e^{-\frac\r\lambda} \Pst(\r)  
\eea
A Yukawa correction has been looked for at various distances
ranging from the millimeter \cite{Adelberger03} 
to the size of planetary orbits \cite{Coy03}.
The accuracy of short range tests has been recently improved, as dedicated experiments were pushed to smaller distances \cite{Chiaverini}, while accounting for Casimir forces 
which become dominant at such ranges \cite{Chen04}.
On the other hand, long range tests of the Newton law, 
performed by monitoring the motions of planets or probes in the solar system,
also agree with GR with good accuracy, for ranges of the order 
of the Earth-Moon \cite{Williams96} to Sun-Mars distances 
\cite{Hellings83,Reasenberg79,Kolosnitsyn03}. 
But these results still show \cite{JR04,Coy03} that windows remain open for violations of the standard form of Newton force law at short ranges, below the millimeter, as well as
long ones, of the order of or larger than the size of the solar system.

 Thus, although they confirm the metric character
of gravitation 
and provide strong evidence in favor of a metric theory lying very close to
GR, gravity tests still leave room for alternative metric theories,
at ranges which have been well tested.
They furthermore  point  to a potential
scale dependence of violations of GR.
Such extensions of GR may in fact arise naturally, in particular when 
effects of radiative corrections are taken into account \cite{tHooft,Fradkin,JR95,Reuter,Hamber07}.

\section{Metric extensions of GR}

After the preceding discussion, we focus our attention on metric theories of
gravitation while releasing the assumption that the solar system
is exactly described by the metric calculated from GR. 
In other words, we consider metric extensions of GR, that is to say
metric theories of gravitation lying in the vicinity of GR. 

Such extensions are more easily explained in their linearized approximation
where the gravitation field is represented as a small perturbation 
$\h_{\mu\nu}$ of Minkowski metric $\eta_{\mu\nu}$ 
\bea
&&\g_\munu = \eta_\munu + \h_\munu \nonumber\\
&&\eta_\munu = {\rm diag}(1, -1, -1, -1) 
\quad,\quad \left\vert \h_\munu \right\vert \ll 1 
\eea
In the linearized approach, the field $\h_{\mu\nu}$ may equivalently be written as a function of position $x$
in spacetime or of a wavevector $k$ in Fourier space, so that Riemann, Ricci, scalar and Einstein curvatures simply read in momentum representation and 
at first order in $\h_{\mu\nu}$
\bea
&&\h_{\mu\nu}(\x) \equiv \int {d^4\k \over (2\pi)^4}e^{-i\k\x} \h_{\mu\nu}[\k]
\nonumber\\
\label{curvatures}
&&\R_{\lambda\mu\nu\rho} = \frac{
\k_\lambda\k_\nu \h_{\mu\rho} - \k_\lambda\k_\rho \h_{\mu\nu}
- \k_\mu\k_\nu \h_{\lambda\rho} + \k_\mu\k_\rho \h_{\lambda\nu}}2\nonumber\\
&&\R_{\mu\nu} = {\R^\lambda}_{\mu\lambda\nu}\quad, \quad \R = {\R^\mu}_\mu\quad,\quad
\E_{\mu\nu} = \R_{\mu\nu} -\eta_{\mu\nu}{\R\over2}
\eea
Then, gravitation equations which generalize Einstein-Hilbert equations (\ref{GR_gravitation_law}) may be
written under the form of a linear response relation between Einstein curvature and 
the energy-momentum tensor \cite{JR95} 
\bea
\label{general_gravitation_law}
\E_\munu [k] = \chi_{\mu\nu\lambda\rho} [k] ~\T^{\lambda\rho} [k] 
\quad,\quad k^\mu \chi_{\mu\nu\lambda\rho} [k] = 0
\eea
GR (\ref{GR_gravitation_law}) corresponds  to a simple local expression for the linear response function \cite{Einstein15,Hilbert}.
Generalized  linear response equations (\ref{general_gravitation_law}) may arise for example from radiative corrections of
GR through the coupling of gravity to quantum fields \cite{tHooft,Fradkin,JR95,Reuter,Hamber07}. 
Then, as gravitation couples to fields with different conformal weights, 
the resulting gravitation coupling differs in the corresponding sectors.
The integral equation (\ref{general_gravitation_law})
takes a simpler form when the energy-momentum tensor 
of the source is static and pointlike ($\M$ is a mass localized at the origin) 
\bea
\label{static_point_source}
\T_{\mu\nu}= \delta_{\mu 0} \delta_{\nu 0} \T_{00}, \qquad \T_{00} =\M\c^2\delta(\k_0)
\eea
In their linearized approximation, extended gravitation equations (\ref{general_gravitation_law}) are then solved for the static point-like source (\ref{static_point_source}), using projectors on the two sectors of
different conformal weights
\bea
&&\E_{\mu\nu}= \E^{(0)}_{\mu\nu} + \E^{(1)}_{\mu\nu}, \qquad \pro_{\mu\nu}\equiv \eta_{\mu\nu} -{\k_\mu \k_\nu\over k^2}\nonumber\\
&&\E^{(0)}_{\mu\nu} = 
\lbrace \pro _{\mu}^{0}\pro_{\nu}^{0}-{\pro_{\mu\nu}\pro^{00}\over3}\rbrace
\,
\frac{8\pi \G^{(0)}}{\c^{4}}\T_{00}, \qquad 
\E^{(1)}_{\mu\nu} =  \frac{\pro _{\mu\nu}\pro ^{00}}{3}
\frac{8\pi \G^{(1)}}{\c^{4}}\T_{00}  \nonumber\\
&&\G^{(0)}= \G_N  + \delta\G^{(0)}, \qquad
\G^{(1)}= \G_N  + \delta\G^{(1)}
\eea
The generalized gravitation equations may then be written in terms of two gravitation 
constants which differ in the two sectors and depend on momentum or length scale.
Two running coupling constants $\G^{(0)}$ and $\G^{(1)}$ thus 
replace the unique Newton gravitation constant $\G_N$, introducing a dependence on the 
conformal weight as well as on the length scale \cite{JR05mpl,JR05cqg}.

Generally, as we restrict our attention to extensions lying in the vicinity of GR,
deviations from GR can be considered as small anomalous terms and calculated
perturbatively. As GR corresponds to Einstein
 curvatures vanishing everywhere except 
on gravity sources, that is the Sun in the solar system, this assumption allows for simplifications. 
For instance from Bianchi identities,
transversality takes a simple form and Einstein tensor can still be
 decomposed on the sectors of traceless and traced tensors 
as in the linearized approximation \cite{JR05mpl,JR05cqg}
(these expressions however preserve their full non linear dependence in Newton gravitation constant $\G_N$)
\bea
\label{perturbed_GR}
&&\E = \stand{\E} + \delta \E, \qquad
\stand{\E} = 0 \qquad {\rm{where}} \qquad \T \equiv 0\nonumber\\ 
\label{Einstein_curvature}
&&\delta\E = \delta \E^{(0)}+\delta \E^{(1)}
\eea
The information contained in the anomalous running coupling constants $\delta\G^{(0)}$ and 
$\delta\G^{(1)}$ is equivalently encoded in the spatial variations of Einstein curvature $\delta \E^{(0)}$ and $\delta \E^{(1)}$ 
of a solution of gravitation equations, or also, of  two independent  
components (for instance in spherical coordinates, $\delta\E^0_0$ and $\delta\E^\r_\r$). 

We begin with the solution of GR gravitation equations (\ref{GR_gravitation_law}) for a static pointlike source (\ref{static_point_source}), which we write in 
isotropic coordinates (\ref{isotropic_metric})
\bea
\label{metric_st}
&&\stand{\g_{00}} = \left({1+\Pst/2\over1-\Pst/2}\right)^2,\qquad
\stand{\g_{\r\r}} = - (1-\Pst/2)^4, \qquad \Pst = -{\G_N\M\over\c^2\r}
\eea
Metric extensions of GR may then be
considered as perturbations of the standard solution  (\ref{metric_st}) 
\bea
\label{perturbed_metic}
\g_{00} =\stand{\g_{00}}+ \delta \g_{00}, \qquad
\g_{\r\r} =\stand{\g_{\r\r}} + \delta \g_{\r\r}
\eea
The perturbed metric components (\ref{perturbed_metic}) should then
correspond to the anomalous Einstein curvatures  (\ref{perturbed_GR})
derived from the modified 
gravitation equations.
Although not knowing the precise form of these equations, one may nonetheless 
write a general form for their solutions. In particular, the two independent components of Einstein curvature characterizing the solution may be chosen so that they lead to simple expressions for the metric components
\bea
\label{anomalous_Einstein_curvature}
&&\delta \E^0_0 \equiv 2 \u^4(\delta \Phi_N - \delta \Phi_P)^{\prime\prime}, \qquad 
\u \equiv {1\over\r(-\stand{\g_{\r\r}})^{1\over2}}, \quad ()^\prime\equiv{d\over d\u}, \qquad 
\stand{\g_{00}} \equiv 1 -{2\G_N\M\over\c^2}\u\nonumber\\
&&\delta \E^\r_\r \equiv 2\u^3\delta \Phi_P^{\prime}
\eea
The metric lying in the vicinity of GR and having Einstein curvatures
as given by equations (\ref{anomalous_Einstein_curvature}) may then be 
expressed in terms of two anomalous potentials  \cite{JR06cqg}
\bea
\label{Eddington_solution}
{\delta\g_{00}\over\stand{\g_{00}}} &=& 2\int {(\delta\PN-\delta\PP)^\prime\over\stand{\g_{00}}^2}d\u
+ 2\int{\delta\PP^\prime \over\stand{\g_{00}} }d\u
+{\stand{\g_{00}}-1\over\stand{\g_{00}}^{1\over2}}
\int{(\delta\PN-\delta\PP)^\prime\over\stand{\g_{00}}^{3\over2}}d\u\nonumber\\
{\delta\g_{\r\r}\over\stand{\g_{\r\r}}} &=& -2\stand{\g_{00}}^{1\over2}
\int {(\delta\PN-\delta\PP)^\prime\over\stand{\g_{00}}^{3\over2}}d\u
\eea
At this stage, it is worth comparing generalizations obtained in this manner with the 
widely used PPN Ansatz \cite{WillNordtvedt72}.
This PPN Ansatz may be written as a particular case of the more general 
extension presented here
\bea
\label{PPN_case}
&&\delta\Phi_N = (\beta-1)\Pst^2 + O(\Pst^3), \qquad 
\delta\Phi_P = -(\gamma-1)\Pst + O(\Pst^2)\nonumber\\
&&\delta\E^0_0 = {1\over\r^2} O( \Pst^2),\qquad
\delta\E^\r_\r = {1\over\r^2} \left(2 (\gamma-1) \Pst +O( \Pst^2) \right)\qquad \mathrm{[PPN]}
\eea
One remarks that the PPN Ansatz corresponds to non zero values for Einstein curvature $\delta\E^\r_\r$ apart from the source,
at first order in $\G\M$, but not for $\delta\E^0_0$ which vanishes at the same order. But, as seen later, the PPN Ansatz does not have the ability to account for the Pioneer anomaly, which is associated to a non vanishing curvature $\delta\E^0_0$, with moreover a  scale dependence which differs from that described by the PPN Ansatz.

\section{Phenomenological consequences}

We now discuss some phenomenological consequences of the metric extensions of
GR. First, we show that they lead to effects on Doppler and ranging measurements which take the form of Pioneer like anomalies.
To describe such observations, one must introduce the time delay function, defined as the time $\td$ taken by a light-like signal to propagate from a spatial position $\xbx$ to another one $\xby$ (representing the coordinates of two different points)
\bea
&&\td(\xbx, \xby), \qquad \xb_a \equiv \r_a (\sin\th_a\cos\ang_a,\sin\th_a\sin\ang_a,\cos\th_a),
\qquad  a=1,2
\eea
The observable used in ranging corresponds to half the time elapsed on Earth between the emission time and the reception time of a signal exchanged between the station on Earth and the probe.
Moreover, as the Pioneer probes were followed by Doppler velocimetry,  their monitoring is more appropriately discussed in terms of the 
Doppler acceleration $\ad$, \cite{Anderson02} which is just the second time derivative of the round trip time delay 
\bea
\label{ranging_time}
&&\Delta \t \equiv  { \td(\xbxm, \xby) +  \td(\xbxp, \xby) \over2}, \qquad
\frac{\ad}{\c}\equiv \frac{d^2\Delta \t}{d\t^2}
\eea
In the case of a static isotropic
space-time the time delay function only depends on three real variables, which can be chosen as the radial coordinates  
$\rx$ and $\ry$ of the two points and the angle $\angxy$ between them as seen from the gravity source.
Its exact form is obtained  by solving  Hamilton-Jacobi
 equation for a light ray
\cite{JR05cqg,JR06cqg}
\bea
\label{time_delay}
&&\c\td(\rx, \ry, \angxy) \equiv \int_\rx^\ry {-{\g_{rr}\over \g_{00}}(\r)
d\r \over \sqrt{-{ \g_{rr}\over \g_{00}}(\r) - {\ri^2\over\r^2}}}
\quad,\quad
\angxy = \int_\rx^{\ry} {\ri d \r/\r^2 \over
\sqrt{-{\g_{rr}\over\g_{00}}(\r) - {\ri^2\over\r^2}}}\nonumber\\
&&\cos\angxy\equiv\cos\th_1\cos\th_2+\sin\th_1\sin\th_2\cos\left(\ang_2-\ang_1\right) 
\eea
Perturbations of GR metric then entail perturbations of the time delay function $\delta\td$ and of the probe trajectory $\delta \xby$ (Earth motion $\xbx$ is assumed to remain fixed)
 which are used for determining the round trip
time delay  (\ref{ranging_time}). Both perturbations affect the observed Doppler acceleration (\ref{ranging_time}), so that when comparing the latter with the result predicted by GR, the residual difference does not vanish and may be interpreted  as an acceleration anomaly affecting the Doppler observable \cite{JR06CQG}
\bea
\label{accelearation_anomaly}
&&\td \equiv \stand{\td} +\delta\td, \qquad \xby \equiv \stand{\xby} +\delta\xby\nonumber\\
&&\ad \equiv \stand{\ad} +\delta\ad
\eea
For an extension in the neighborhood of GR,
the acceleration anomaly may be obtained in terms of the two
metric components which characterize the perturbed static isotropic metric
(\ref{perturbed_metic},\ref{Eddington_solution}).
The expression obtained for the acceleration anomaly in the general case \cite{JR06CQG} simplifies in the practical case of the Pioneer probes, as several contributions may be neglected in a first step.
First,  the GR metric components differ from  1 for $\stand{\g_{00}}$ and from  -1
for $\stand{\g_{\r\r}}$ by the Newton potential, which at the distances explored by Pioneer probes is smaller than  $10^{-9}$. Corrections induced by 
the square of the ratio of the probe velocity to light velocity are of the same order ($\sim 10^{-9}$). Then, the impact parameter of the signal
$\stand{\ri} (\sim \rx $) is at least 20 times smaller than the probe distance $\stand{\ry}$. 
Finally, the probes follow almost linear trajectories. 
Neglecting all the corresponding correcting terms,
 the acceleration anomaly reduces to the sum of two main contributions 
\cite{JR06CQG}
\bea
\label{simplified_Pioneer_anomaly}
&&\delta\ad \simeq \delta\ad _\sec + \delta\ad_\ann\nonumber\\
&&\delta\ad _\sec \simeq -{\c^2\over2} \partial_\r(\delta\g_{00})
+\stand{\rydd}\left\lbrace{\delta(\g_{00}\g_{\r\r})\over2} -\delta\g_{00}\right\rbrace
-{\c^2\over2}\partial_\r^2\stand{\g_{00}}\delta\ry \nonumber\\
&&\delta\ad _\ann \simeq {d\over d\t}\left\lbrace\stand{\dang}\delta\ri\right\rbrace 
\eea
The term $\delta\ad _\sec$ contains secular contributions proportional to metric anomalies and depending on the probe accceleration, as well as to a range ambiguity $\delta\ry$.
As unfortunately no range capabilities were available for the Pioneer probes, the latter should be determined by fitting the probe trajectory with the Doppler signals using a given model.
The term $\delta\ad _\ann$ is a modulated contribution, proportional to the relative angular velocity between the probe and Earth and to an anomaly of the impact parameter $\delta \ri$.
 Using a simplest form for the impact parameter $\stand{\ri}$,
 modulated by Earth rotation, and taking as reference 
the acceleration on Earth due to the 
Sun gravitation field, the secular and modulated anomalies may be expressed in terms of the potential anomalies in the two different sectors \cite{JR06CQG}
\bea
\label{oversimplified_Pioneer_anomaly}
&&\delta\ad _\sec \simeq -\c^2 \partial_\r\delta\PN (\ry) 
+\aE {\rx^2\over\ry^2} \left\lbrace { 2\delta\ry \over\ry} + \delta\PP (\ry) + 2\delta\PN (\ry) \right\rbrace
\nonumber\\
&&\delta\ad_\ann \simeq -\annfactor\cos\left(2\Ox(\t-\tconj)\right) 
-{d\annfactor\over2\Ox d\t} \sin\left(2\Ox(\t-\tconj)\right) 
\nonumber\\
&&\annfactor \equiv \aE{\rx\over\ry} \left\lbrace
{\delta\ry\over\ry} - \ry \int_\rx^{\ry} (\delta\PP-2\delta\PN) {d\r\over\r^2} 
\right\rbrace\nonumber\\
&&\aE\equiv  {\G_N \M\over\rx^2} = \Ox^2\rx \simeq 6\times10^{-3}\mathrm{m\,s}^{-2}
\eea
The time $\tconj$ corresponds to the closest approach between Earth and probe.
The anomaly $\delta\ad_\ann$ depends on an amplitude $\annfactor$ 
which is determined by the range ambiguity and the two potential anomalies, and is modulated at twice the orbital frequency.

The secular and modulated anomalies involve 
different linear superpositions of the range ambiguity and
 the potentials (\ref{oversimplified_Pioneer_anomaly}). 
Although the modulated anomaly seems to be
 larger than the secular one by a factor $\ry/\rx$, 
this turns out not to be the case in practice  \cite{Anderson02}.
Indeed, as the data analysis process 
is based on the \textit{a priori} assumption that there
is no  modulated anomaly in the physical signal,
the best trajectory fitting the data
tends to produce a quasi null value for the modulated anomaly, i.e.
a nearly perfect 
compensation of its different contributions.
Such a compensation has to be effective
within a fraction of the order of $\rx/\ry$ 
but cannot hold over a long period of time, due to
 the different range dependences of the compensating terms 
and also  to the maneuvers which change the constants
of motion, hence the range ambiguity.
A precise estimation of the modulated  anomaly would then require a complete solution of the
motion including a detailed description of the maneuvers.
The latter being unavailable, we shall  use the previous description
to obtain a rough quantitative estimate of the secular anomaly.
Inserting for the range ambiguity in (\ref{oversimplified_Pioneer_anomaly}) the value minimizing the modulated anomaly  and keeping leading contributions only,
one obtains a general expression for the secular anomaly which only depends on the two potential anomalies
\bea
\label{Pioneer_anomaly}
&&\delta\ad _\sec \simeq -\c^2 \partial_\r\delta\PN (\ry) +
\aE \rx^2 \left\lbrace {2\over\ry} \int_\rx^{\ry} \delta\PP(\r) {d\r\over\r^2}
  + {  \delta\PP(\ry)\over\ry^2}
\right\rbrace 
\eea
One should note on the final form of the anomalous acceleration (\ref{Pioneer_anomaly})
its non linear dependence on the two gravitation potentials, as it involves a contribution which is the product of the standard Newton acceleration (first sector) by  the potential in the second sector.
In particular, this result differs significantly from those obtained from
preliminary studies using the same framework \cite{JR05cqg,JR06cqg} but neglecting  the modulated anomalies, which in fact appear to play a crucial role.
  
According to the expression (\ref{Pioneer_anomaly}) for the secular anomaly, a constant anomalous acceleration ($\delta\ad _\sec  = -\aP/\c^2 \equiv -1/\H$), as observed in the distance ranges of the Pioneer 10/11 probe, may be obtained in several ways.
First, assuming a  perturbation limited to the sector of Newton potential, one obtains
that the latter must depend linearly on the heliocentric distance  
\bea
\label{model_N}
\delta \Phi_N \simeq \frac {\r}{\H}
\eea
However, the effects of a metric keeping the same form (\ref{model_N}) between Earth and Mars could not have escaped detection in tests 
performed with martian probes \cite{Reasenberg79,Anderson02,JR04}. 
The perturbation (\ref{model_N})  would produce a change of planet 
orbital radii, leading to range variations of the order of $50km$ to $\sim$100km,
while the Viking data constrain these measurements to agree with standard expectations
at a level of $\sim$100m to $\sim$150m. One should further note that 
the effect of the Shapiro time delay in the range evaluation has here a negligible influence
\cite{Moffat06}. A remaining possibility is that the perturbation (\ref{model_N})
be cut off at heliocentric distances corresponding to inner planets 
\cite{Brownstein06}, a possibility which should then be confronted with 
available data on the ephemerids of outer planets.

If on the other hand the perturbation is limited to the second sector,
a constant anomalous acceleration is obtained from a quadratic dependence of the
potential $\delta\PP$ on the heliocentric distance 
\bea
\label{model_P}
\delta \PP \simeq -{\c^2\over 3\G_N \M}\frac{r^2}{\H}
\eea
The metric perturbation (\ref{model_P}) also has an influence
on the already discussed range measurements on martian probes.
If the dependence (\ref{model_P}) were extended without modification to the inner part of the solar system, the Shapiro effect would lead to  range variations which stand in slight conflict with Viking data.
But this discrepancy  may be cured by mildly cutting off the simple dependence (\ref{model_P}) 
at the orbital radii of inner planets, without affecting predictions 
made for the Pioneer probes. 
In any way, a precise analysis of the anomalous acceleration (\ref{Pioneer_anomaly}) should consider the most general form depending on the potentials in both sectors (such as a linear combination of the two simple models (\ref{model_N})  and (\ref{model_P})).
The evaluation of correlated effects of anomalies in the two sectors will
be necessary to confront observations with theory in a significant way.

We now discuss effects of the extended metric in the inner part of the solar
system. An important success of GR concerns its agreement with planet
ephemerids. In the case of metric extensions of GR, and
for the nearly circular orbits followed by planets, the 
anomalous precession
may be given as an expansion in the orbit eccentricity $\ep$.
The result, derived and discussed in \cite{JR06cqg}, is just given here 
\bea
\label{perihelion_constraint_circular}
{\delta\Delta \prec\over 2\pi} &\simeq& 
\u\left(\u\delta\Phi_P\right)^{\prime\prime} 
-{\c^2 \u\over2\G_N \M}\delta\Phi_N^{\prime\prime} 
+ {\ep^2 \u^2\over8}\left(\left(\u^2\delta\Phi_P^{\prime\prime}
+\u\delta\Phi_P^\prime\right)^{\prime\prime} 
-{\c^2\u\over2\G_N \M}\delta\Phi_N^{\prime\prime\prime\prime} 
\right) + \ldots
\eea
The well known expression of the anomalous perihelion precession 
in the PPN formalism \cite{Will01} is recovered by inserting in (\ref{perihelion_constraint_circular}) 
the particular expressions of $\delta\PN$ and $\delta\PP$ for the PPN metric
(\ref{PPN_case}).
Keeping only the leading term, one obtains the anomalous perihelion precession in
terms of anomalous parts of Eddington parameters $\beta$ and $\gamma$
\bea
\label{PPN_precession}
\left({\delta\Delta \prec\over2\pi}\right)^{PPN} &\simeq& \left(2(\gamma-1)-(\beta - 1)\right) 
{\G\M\over\c^2}\u
\eea
The more general expression (\ref{perihelion_constraint_circular}) may thus be thought of as a generalization of the anomalous PPN parameters $\beta-1$ and 
$\gamma-1$ to functions $(\delta\PN)^{\prime\prime}$ and $(\u\delta\Phi_P)^{\prime\prime}$ which may depend
on the heliocentric distance. Expression (\ref{perihelion_constraint_circular}) for perihelion precession anomalies
 may then be used to constrain the potential anomalies 
for ranges around a few AU where planet ephemerids are precisely known.

Finally, we come to the discussion of light deflection experiments by the Sun
gravitation field, which amount to explore 
the potentials $\PN$ and $\PP$ in the Sun vicinity.
To this aim, we first rewrite the definition of the anomalous part $\delta\gamma$
of the Eddington parameter.
We decompose the relevant combination of the two potentials $2 \delta\Phi_N -\delta \Phi_P$ and separate the part singular at $\r=0$ from the regular part describing the long range effect
\bea
\label{modified_Eddington}
 2 \delta\Phi_N(\r) -\delta \Phi_P(\r)  \equiv -{\G_0 \M \over \c^2 \r}
 + {\M \over \c^2 } \r \zg_0 (\r)
\eea
The coefficient $\G_0$ of the singular part in (\ref{modified_Eddington}) corresponds to the standard
PPN modification of Eddington parameter $\gamma$.
One specific feature characterizing the extended metric is then the  
dependence of the anomalous part of Eddington parameter $\delta\gamma$ on the impact parameter $\ri$ (of the electromagnetic signal deflected by the Sun gravitation field), which is determined by the regular part of the anomalous potential
\bea
\delta \gamma(\rho) = -\frac{\G_0}{\G_N} +
\frac{\zeta_0(\rho) \rho^2}{2\G_N} 
\eea
This dependence, which allows one to reveal the long range variation of the 
potential $\PP$,  can be looked for  by keeping trace
of the whole functional dependence of the anomalous deflection angle
during an occultation by the Sun 
\bea
\delta \Delta\theta \simeq  
- {\G_N \M\over c^2}{\partial \over \partial \rho}
\left( \delta \gamma(\rho){\rm{ln}}{4 \r_1 \r_2 \over \rho^2}\right) 
\eea
For experiments within the solar system, $\rx$ has to be replaced by the Sun-Earth distance
and $\ry$ by the Sun-probe distance. 
When the emitter is a source far outside the solar system,
$\ry$ has to be replaced by a cutoff distance.
Evidence of a variation of $\gamma$ would provide a direct
validation of the extended framework. 
This aim might be attainable through a reanalysis of existing data
as those of the Cassini experiment \cite{Bertotti03}.
In the future, it can also be reached by high accuracy Eddington tests (see
for example the LATOR project \cite{Lator}) or global mapping of deflection
over the sky (see for example the GAIA project \cite{Gaia}). 
The former test would lead to an accuracy on the measurement of $\gamma$ 
sufficient to see the effect if its order of magnitude corresponds to
that of the Pioneer anomaly. The latter test
would allow one to reconstruct the dependence of $\gamma$ versus the impact
parameter $\ri$ and then, the dependence of $\PP$ versus $\r$.

To summarize, the previous discussion of the effects associated with 
 metric extensions of GR
 comforts the possibility to modify GR at large scales, while preserving the
 compatibility with existing classical gravity tests and allowing
to include the Pioneer probes test. It also indicates that metric extensions 
could lead to new and observable effects at distance scales typical of the solar system.
This provides further  motivation for performing new experiments in the solar system in order to improve our knowledge of gravity laws, in particular of
their scale dependence.

\section{Acknowledgements}

We would like to thank for discussions the members of the ``Deep Space Gravity Explorer'' team
(H.~Dittus \etal) \cite{PAEM05}, of the ``Pioneer Anomaly Investigation Team'' 
(S.G.~Turushev \etal) \cite{Turyshev06}, and of the ``Groupe Anomalie Pioneer''.

\end{document}